\documentclass[10pt,conference,letterpaper,nofonttune]{IEEEtran}
\IEEEoverridecommandlockouts
\usepackage{amsmath,amssymb,amsfonts}
\usepackage{algorithmic}
\usepackage{graphicx}
\usepackage{textcomp}
\usepackage{xcolor}
\usepackage{balance}

\usepackage[numbers,sort&compress]{natbib}
\usepackage{tikz}
\usepackage{tikzscale}
\newif\ifexttikz
\exttikzfalse
\ifexttikz
\else
\usepackage{tikzpagenodes,etoolbox}
\usetikzlibrary{calc}
\usepackage[contents={}]{background}
\AddEverypageHook{%
\ifnumequal{\thepage}{1}{%
    \tikz[remember picture,overlay]{%
        \node[draw,
        minimum width=1.03\textwidth,
        text width=1.02\textwidth,
        font=\footnotesize
        ]
        at ($(current page header area) - (0,5pt)$)
        {%
        This paper has been accepted for publication on International Conference on Machine Learning for Communication and Networking (ICMLCN 2024). This is the author's accepted version of the article. The final version published by IEEE is CP. Robinson, D. Uvaydov, S. D'Oro, and T. Melodia, ``DeepSweep: Parallel and Scalable Spectrum Sensing via Convolutional Neural Networks'' \textit{IEEE ICMLCN 2024 - IEEE International Conference on Machine Learning for Communication and Networking}, Stockholm, Sweden, 2024.
        };
        \node[draw,
        minimum width=1.03\textwidth,
        text width=1.02\textwidth,
        font=\footnotesize
        ]
        at (current page footer area)
        {%
        ©2024 IEEE. Personal use of this material is permitted. Permission from IEEE must be obtained for all other uses, in any current or future media, including reprinting/republishing this material for advertising or promotional purposes, creating new collective works, for resale or redistribution to servers or lists, or reuse of any copyrighted component of this work in other works.
        };
    }%
}{}
}
\fi

\begin{document}

\title{\textit{DeepSweep}: Parallel and Scalable Spectrum Sensing via Convolutional Neural Networks}

\author{\IEEEauthorblockN{Clifton Paul Robinson*, Daniel Uvaydov*, Salvatore D'Oro*, and Tommaso Melodia*}
\IEEEauthorblockA{*Institute for the Wireless Internet of Things, Northeastern University, United States\\
Email: $\{$robinson.c, s.doro, uvaydov.d, t.melodia$\}$@northeastern.edu}

\vspace{-25pt}

\thanks{Research was sponsored by the Army Research Laboratory and was accomplished under Cooperative Agreement Number W911NF-23-2-0014. The views and conclusions contained in this document are those of the authors and should not be interpreted as representing the official policies, either expressed or implied, of the Army Research Laboratory or the U.S. Government. The U.S. Government is authorized to reproduce and distribute reprints for Government purposes notwithstanding any copyright notation herein.}
}


\maketitle

\begin{abstract} 
Spectrum sensing is an essential component of modern wireless networks as it offers a tool to characterize spectrum usage and better utilize it. Deep Learning (DL) has become one of the most used techniques to perform spectrum sensing as they are capable of delivering high accuracy and reliability. However, current techniques suffer from ad-hoc implementations and high complexity, which makes them unsuited for practical deployment on wireless systems where flexibility and fast inference time are necessary to support real-time spectrum sensing. In this paper, we introduce \textit{DeepSweep}, a novel DL-based transceiver design that allows scalable, accurate, and fast spectrum sensing while maintaining a high level of customizability to adapt its design to a broad range of application scenarios and use cases. \textit{DeepSweep} is designed to be seamlessly integrated with well-established transceiver designs and leverages shallow convolutional neural network (CNN) to ``sweep" the spectrum and process captured IQ samples fast and reliably \textit{without} interrupting ongoing demodulation and decoding operations. \textit{DeepSweep} reduces training and inference times by more than 2 times and 10 times respectively, achieves up to 98\% accuracy in locating spectrum activity, and produces outputs in less than 1~ms, thus showing that that \textit{DeepSweep} can be used for a broad range of spectrum sensing applications and scenarios.
\end{abstract}

\begin{IEEEkeywords}
Spectrum sensing, deep learning (DL).
\end{IEEEkeywords}

\section{Introduction}
\label{sec:intro}

\begin{figure}[!t]
    \centering
    \includegraphics[width=\linewidth]{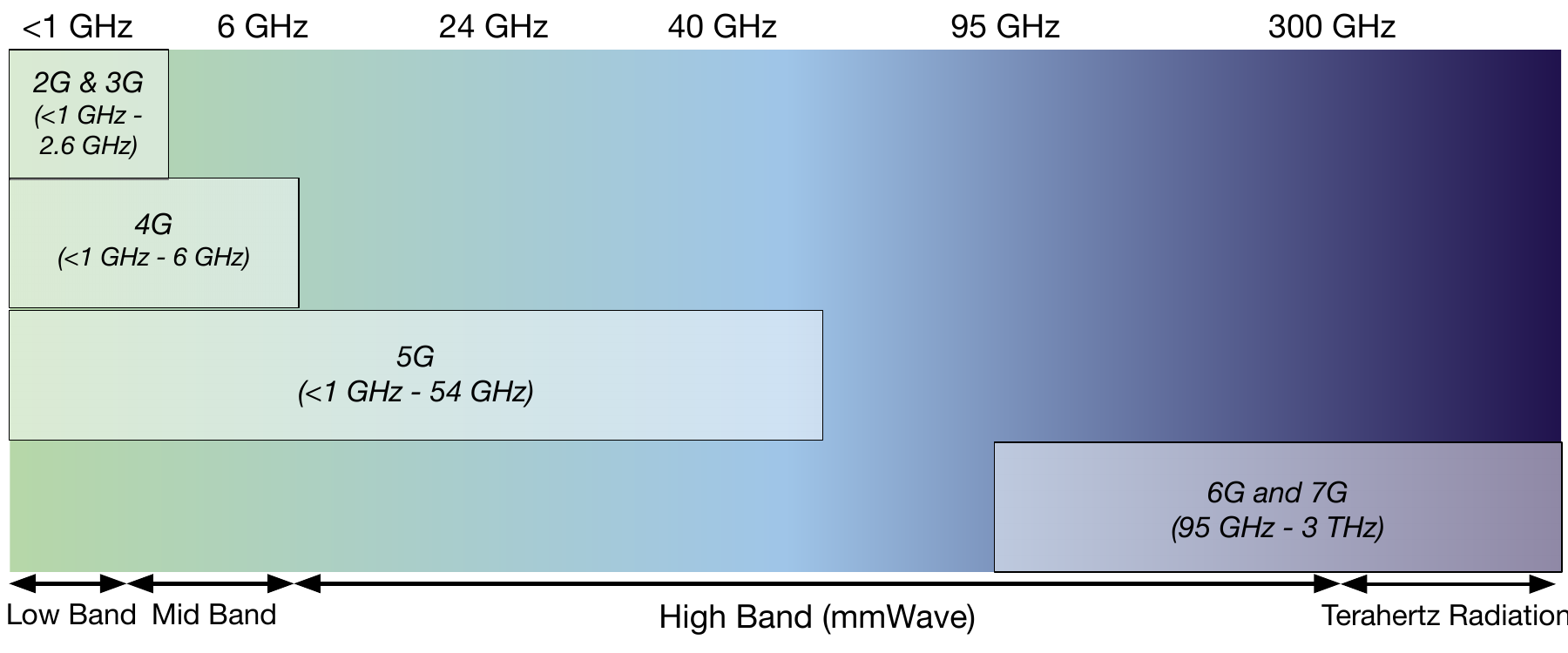}
    \vspace{-0.7cm}
    \caption{A high-level overview of the wireless spectrum, showing how as the technology advances, more of the spectrum is used within our communication.}
    \label{fig:spectrum_g}
\end{figure}

As the number of wireless devices connected to the Internet increases, spectrum scarcity has become more and more evident~\cite{kim2020opportunism}. Traditionally, this scarcity has been mitigated by expanding the portion of the spectrum we utilize. Fig.~\ref{fig:spectrum_g} shows how spectrum has evolved from one cellular generation to another, starting from 2G and 3G which only occupied spectrum frequencies to 2.6~GHz, to 5G which operates at millimeter wave bands and 6G gradually transitioning into the terahertz region of the spectrum. 
On the one side, this approach has the advantage of replenishing spectrum availability and providing the necessary bandwidth to support the ever-increasing demand for high data rates. On the other side, 
this process takes time, as the transition from one ``G" to the other takes approximately a decade, which leaves spectrum availability an open challenge. 

Since utilizing the spectrum is a necessity for any wireless system, the last decade has gravitated around the concept of \textit{spectrum sensing}. This technology consists in monitoring a certain portion of the spectrum to detect the presence of spectrum activity and identify unused portions of the spectrum, i.e., \textit{spectrum holes}, that can be accessed opportunistically~\cite{DeepSense}. One particular example of relevance is that of Citizens Broadband Radio Service (CBRS), which is a 150~MHz spectrum band in the 3.55-3.7~GHz range that the Federal Communications Commission (FCC) has allocated to opportunistic spectrum access. Specifically, cellular users can use the CBRS band in the absence of incumbents (e.g., radar and naval systems), but need to rapidly vacate this band as soon as incumbent transmissions start and are detected via spectrum sensing.    
To be effective, spectrum sensing needs to be both fast and accurate, meaning that the spectrum sensing system should be able to characterize with high accuracy the monitored spectrum and output such characterization within the coherence time of the observation. Moreover, since the portion of spectrum to be monitored keeps increasing from one G to the other, we also need scalable solutions that can deliver fast and accurate sensing capabilities even when monitoring large portions of spectrum.

To meet the above requirements, recently there has been increasing interest in using machine learning (ML), and more specifically, deep learning (DL), as potential tools to deliver spectrum sensing capabilities. This is because these technologies can learn the underlying dynamics of a system directly from data without relying upon a mathematical model of the system-defined a-priori. Thanks to these properties, ML and DL models have been shown to achieve high classification and detection accuracy while being both resilient against stochastic channels and implementable on commercial off-the-shelf (COTS) hardware~\cite{restuccia2021deepfir, zha_deep_2019, arjoune_novel_2020, uvaydov_deepsense_2021, soltani_real-time_2019}.

Although DL can effectively deliver spectrum sensing capabilities, currently available solutions are still not able to deliver at the same time real-time inference, high accuracy, and scalability. Moreover, they generally target specific use cases and hardly apply to different applications, which makes them hard to customize. For example, \cite{GaoDKSS} offers accurate sensing, but combines convolutional neural networks (CNNs) and long short-term memory (LSTM) which slow down inference time. In \cite{bhatti_shared_2021}, authors demonstrate accurate spectrum sensing on COTS hardware but lack scalability and need more than 100ms to classify the spectrum. Similarly, \cite{dl_wsc} presents scalable spectrum sensing solutions at the price of reduced accuracy.

In this paper, we aim to bridge the gap between the above work and practical application by designing and prototyping a system capable of delivering spectrum sensing capabilities that are accurate, fast, and scalable at the same time. To achieve our goal, we propose \textit{DeepSweep}, a novel data-driven transceiver design that allows scalable, accurate, and fast spectrum sensing while maintaining a high level of customizability to adapt its design to a broad range of application scenarios and use cases. 

\textit{DeepSweep} performs spectrum sensing \textit{without} interrupting ongoing demodulation and decoding operations, and its design is based on a spectrum sensing pipeline where incoming waveforms are first converted into the frequency domain and then split into a set of equally sized spectrum chunks. Chunks are then processed in parallel and in batch by a high-resolution DL model that, once computed, combines the individual outcomes of each chunk into a single output that describes spectrum utilization in the entire monitored band. We show that this procedure makes it possible to reduce training and inference times by more than 2$\times$ and 10$\times$ respectively if compared to a model with the same resolution but processing the entire monitored band as a whole. 

\textit{DeepSweep} is trained and tested exclusively on datasets created from signals collected ``in the wild". We demonstrate \textit{DeepSweep} in a real-world scenario where the goal is to detect and locate with high accuracy narrowband interfering signals.  We prototype \textit{DeepSweep} on COTS wireless radios and test it with over-the-air data. Our experiments show that \textit{DeepSweep} can deliver up to 98\% detection accuracy and add an inference latency lower than 1~ms.

\section{Related Work}
\label{sec:relate}

In recent years, spectrum sensing has received much interest from the research community~\cite{ss_survey_1, xie_parallel_2010, liu_advanced_2013, DeepSense, davaslioglu_deepwifi_2021, bhatti_shared_2021, spec_comp, narrow_sense}. 

Recently, DL has gained increasing interest as a solution to solve a variety of problems in the wireless domain, with spectrum sensing being one of the major practical applications~\cite{ss_dl_cr, dl_wsc, pattern_ss_cnn, DeepSense, GaoDKSS, blind_ss}. Being able to learn directly from data, DL-based approaches enable accurate, efficient, and adaptive spectrum sensing solutions that can operate in complex and dynamic environments. 
In \cite{ss_dl_cr}, the authors train their models to detect 8 specific signal types as well as the lack of transmission activities. Similarly, \cite{dl_wsc} focuses on classifying different modulations based on signal-to-noise ratio (SNR) conditions using LSTMs that execute on low-cost sensors, and show that their approach achieves an accuracy level between 75\% and 90\% depending on modulation type and SNR levels.
%
The authors in \cite{DeepSense} propose a wideband spectrum sensing designed to identify spectrum holes with high accuracy and low latency using CNNs embedded in the transceiver chain.

There has also been a shift towards DL-based spectrum sensing with a focus on those cases where labeled datasets are unavailable or incomplete. For example, the authors in \cite{blind_ss} combine three different neural network (NN) architectures (i.e., CNNs, LSTMs, and Fully Connected NNs) to perform blind spectrum sensing. The authors show that their approach outperforms energy detectors even at low SNR levels. \cite{GaoDKSS} takes a different approach by exploiting the underlying structural information of modulated signals to achieve detection with over 99\% accuracy without any prior information about the channel or background noise.

Another application that has witnessed increasing interest is that of spectrum sharing. In this case, spectrum sensing plays a pivotal role in being an enabler to determine which portions of the spectrum should be used and avoid those already being used~\cite{liu_advanced_2013, bhatti_shared_2021}. Previous research has focused on classifying portions of the spectrum and the type of waveforms being transmitted at certain frequencies. For example, \cite{bhatti_shared_2021} uses CNNs to detect the presence of transmission activities including WiFi and LTE. Moreover, other works have focused on using parallel processing to improve the flexibility and latency of trained models for spectrum sensing applications~\cite{liu_advanced_2013}.

Although the above models offer high accuracy, they either result in DL models that are computationally heavy, which eventually results in high inference time (even in the order of several hundreds of milliseconds); or have been tested on simulated data only, leaving their feasibility and applicability to real-world applications unclear. Moreover, most works targeting wideband classification do not deliver high resolution and might fail in detecting narrowband signals and interference. In this work, we try to address the above limitations altogether by developing \textit{DeepSweep}, a lightweight and highly parallelizable spectrum sensing algorithm designed to perform inference in \textit{real time} and with high resolution. Since our goal is to make our approach suitable to practical over-the-air operations, we pursue an approach that is radically different from previous works and train and test \textit{DeepSweep} on data exclusively collected ``in the wild" on a wireless testbed which will be described in Section \ref{subsec:testbed}.

\section{\textit{DeepSweep} System}
\label{sec:frame}

To provide a reliable, fast, and general framework to detect and characterize wideband spectrum portions, we have designed, developed, and prototyped \textit{DeepSweep}, a novel spectrum sensing system that operates at the physical layer by converting received IQ samples into useful information on spectrum usage and occupancy. Fig.~\ref{fig:deepsweep} provides the high-level overview of the main components of \textit{DeepSweep} and their integration with orthogonal frequency-division multiplexing (OFDM) transceiver being one of the most used technologies in wireless systems such as 5G, 4G, and WiFi.

\subsection{Architecture and Procedures}

\begin{figure}[!t]
    \centering
    \includegraphics[width=\linewidth]{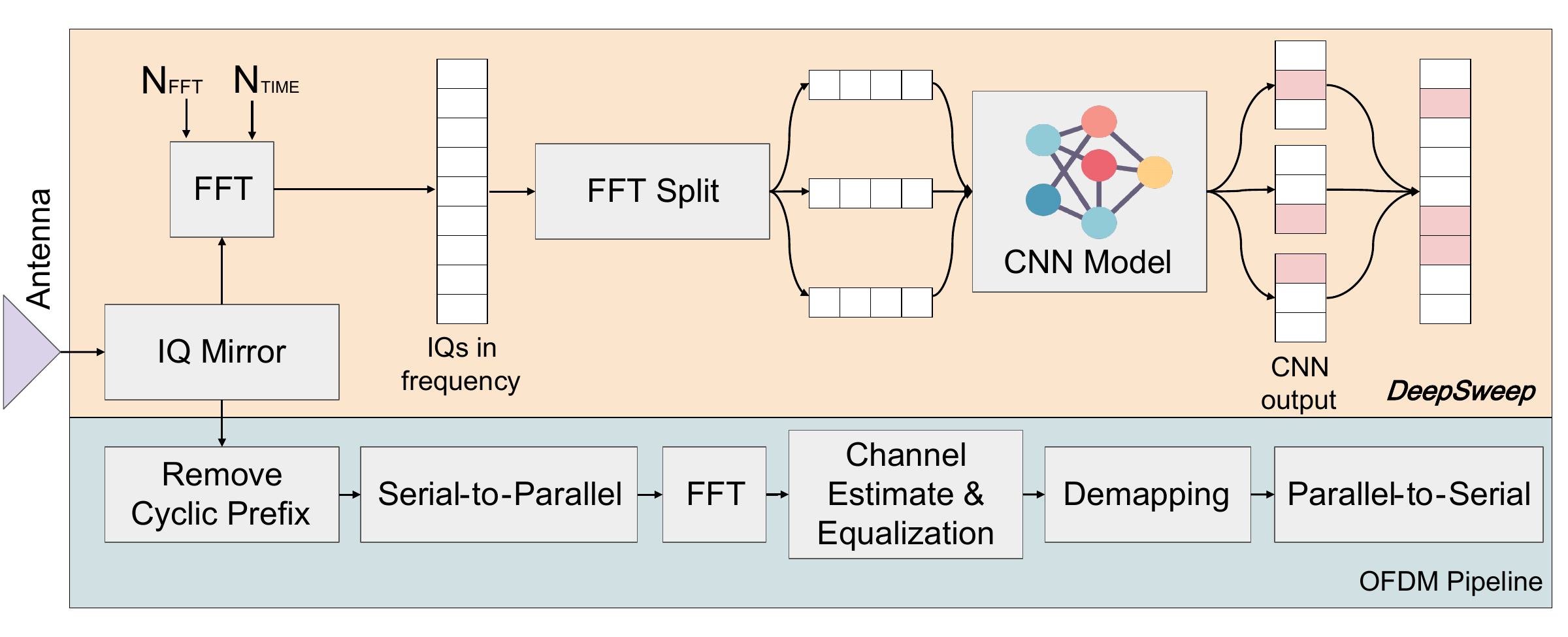}
    \vspace{-0.7cm}
    \caption{The high-level design of \textit{DeepSweep}, showing how the OFDM RX chain and \textit{DeepSweep} work in parallel without interruption.}
    \label{fig:deepsweep}
\end{figure}

As shown in the top part of Fig.~\ref{fig:deepsweep}, \textit{DeepSweep} is constantly running in the background as a software daemon that performs spectrum sensing tasks in parallel to demodulation and decoding tasks. First, IQ samples are captured by the \textit{IQ Mirror} module, whose goal is to duplicate received baseband IQ samples so that one copy is fed to the standard OFDM processing pipeline (Fig.~\ref{fig:deepsweep} (bottom)), and the other copy to the \textit{DeepSweep} modules. The goal of the IQ Mirror is to ensure that IQs are not queued, and \textit{DeepSweep} operations do not interrupt the normal decoding and demodulation procedures (which would instead result in loss of data and synchronization). 

The time-domain mirrored baseband IQs are then processed by \textit{DeepSweep}'s \textit{Spectrum Sweeping} module. Specifically, captured IQs are projected into the frequency domain via a Fast Fourier Transform (FFT) operation that converts $N_{\mathrm{TIME}}$ time-domain IQ samples into $N_{\mathrm{FFT}}$ IQ samples in the frequency domain. Both parameters are customizable and can be used to regulate the sensing resolution of \textit{DeepSweep} (i.e., $N_{\mathrm{FFT}}$) as well as the observation temporal window (i.e., $N_{\mathrm{TIME}}$).

The $N_{\mathrm{FFT}}$ frequency-domain IQ samples are then split by an \textit{FFT Partitioner} into $G$ groups, each of size $N_{\mathrm{FFT}}/G$ IQ samples. At this point, the $G$ groups are fed to the \textit{Spectrum Sensing CNN} as a batch and processed in parallel to produce $G$ distinct outputs. These are then combined into a single output by the \textit{Output Processor}. 

\textit{DeepSweep} offers a general design where the Spectrum Sensing CNN is not tied to any specific problem or task as it can be reprogrammed to host any CNNs. Moreover, since the FFT module is tunable, it offers a reconfigurable platform to vary both the resolution in both frequency (i.e., $N_{\mathrm{FFT}}$) and time (i.e., $N_{\mathrm{TIME}}$) domains. 

\subsection{Spectrum Sweeping}

The major advantage of \textit{DeepSweep} lies in its spectrum sweeping capability which makes it possible to support high-resolution spectrum sensing with low complexity and fast inference time. 

The most common approach in the literature consists in training a CNN that processes all of the $N_{\mathrm{FFT}}$ IQs at the same time and performs spectrum sensing on the entire monitored band, say $W$. This can indeed offer high accuracy but, as we will demonstrate in Section \ref{sec:results}, might result in high inference latency that could only be mitigated by reducing the resolution of the CNN. The underlying reason for this behavior is that the higher the complexity of the input (i.e., $N_{\mathrm{FFT}}$), the more complex the CNN needs to be (e.g., more convolutional layers, deeper architectures, more neurons) to properly capture all of the features of the wireless signals, and the higher is the inference latency (due to the higher number of layers and neurons of the CNN). 

In \textit{DeepSweep}, we follow a radically different approach. Rather than training and deploying a spectrum sensing CNN that processes $N_{\mathrm{FFT}}$ IQ samples at once, we use a CNN that processes a batch of $G$ inputs, each consisting of $N_{\mathrm{FFT}}/G$ IQ samples. The individual $G$ outputs are then combined into a single output that represents the unified spectrum sensing outcome of the CNN. The advantage of this approach is that we can achieve high-resolution spectrum sensing capabilities without necessarily increasing the complexity of the CNN. Instead, we can use a simpler and shallower CNN, and leverage parallel processing of multiple chunks of the same input to characterize the spectrum to maintain the complexity of the spectrum sensing task low. Indeed, (i) we can increase the resolution ($N_{\mathrm{FFT}}$) of the spectrum sensing task and simply increase the number of groups $G$ that are fed to the CNN; (ii) we can use the same CNN independently of the width of the observed band $W$ and for varying sizes of $N_{\mathrm{FFT}}$ without any need to retrain the CNN, which offers a more scalable and portable solution if compared to the traditional approach of retraining CNNs and enables scalable wideband sensing; (iii) the inference and training time of a shallow CNN with input size $N_{\mathrm{FFT}}/G$ can be significantly lower than a deeper one that is trained to process inputs of size $N_{\mathrm{FFT}}$.

\begin{figure}[!t]
    \centering
    \includegraphics[width=0.95\linewidth]{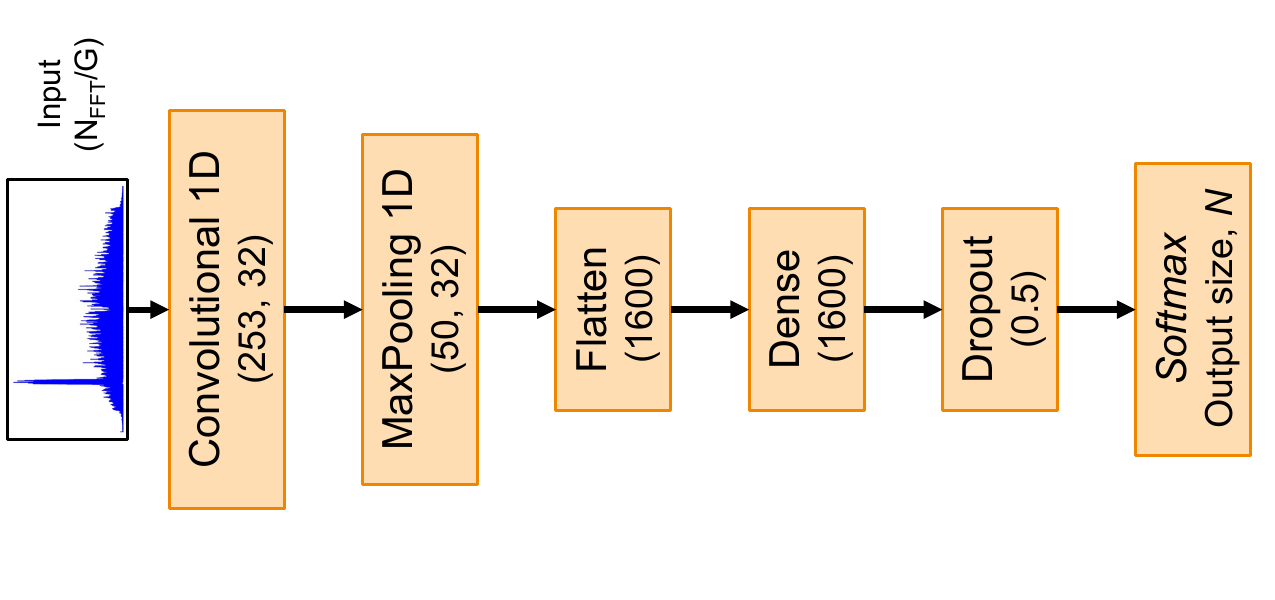}
    \vspace{-0.5cm}
    \caption{The architecture of the Spectrum Sensing CNN.}
    \label{fig:jam_cnn}
\end{figure}

\subsection{Spectrum Sensing CNN Architecture \& Training}
\label{subsec:cnn}

The main goal of \textit{DeepSweep} is to offer a scalable and lightweight spectrum sensing system that can operate in real-world deployments and deliver real-time inference. 

CNNs have been demonstrated to be highly accurate in performing spectrum sensing~\cite{cnn_acc} by only using IQ samples as inputs, especially when considering IQ samples in the frequency domain~\cite{ffts}. We follow the same approach to design \textit{DeepSweep}'s spectrum sensing CNN. 

Although \textit{DeepSweep} is general and supports the use of different CNN architecture, in this paper we have used a simplified version of single convolutional model (conv-to-maxpool pair).
Fig.~\ref{fig:jam_cnn} shows the architecture of the one-dimensional (1D) CNN used by \textit{DeepSweep}. 
Our model takes as input $N_{\mathrm{FFT}}$ IQ samples in the frequency domain. The input is then processed by a single 1D convolutional (Conv1D) layer, followed by a 1D maximum pooling (MaxPool1D) layer with a stride of 4. 
The latter is then followed by Flatten and Dense layers. The penultimate layer consists of a Dropout layer to help combat overfitting, then is followed by a Softmax layer to produce the final output of the CNN.

The benefits of this architecture are at least twofold. First, it only uses a convolutional layer, which greatly reduces the computational complexity and inference time of the CNN. Second, it strikes the right balance between generalization (i.e., it hardly overfits) and accuracy. Those benefits will be illustrated via experimental results in Section \ref{sec:results}.


The CNN is trained using the Adam optimizer \cite{adam}, we use the categorical cross-entropy loss function and a learning rate $\gamma=0.01$ with early stopping.  


\subsection{Use Case: detecting narrowband interference}
\label{subsec:usecase}

Although \textit{DeepSweep} is general and can tackle a variety of applications requiring accurate, fast, and precise spectrum sensing, due to the lack of space in this paper we focus on a specific use case that matches the above requirements. Specifically, we consider the case of narrowband interference detection. 
Narrowband interference can affect pilot symbols commonly used in most wireless standards and is necessary to complete successful decoding procedures. An example of relevance is that of voluntary interference, or pilot jamming~\cite{pilot}. We train \textit{DeepSweep} to focus on detecting such signals.
%

\section{Experimental setup and dataset}
\label{sec:datasets}

In this section, we describe both the testbed and the datasets we use to train and test \textit{DeepSweep}.

\subsection{Over-the-air Prototype}
\label{subsec:testbed}

\begin{figure}[!t]
    \centering
    \includegraphics[width=\linewidth]{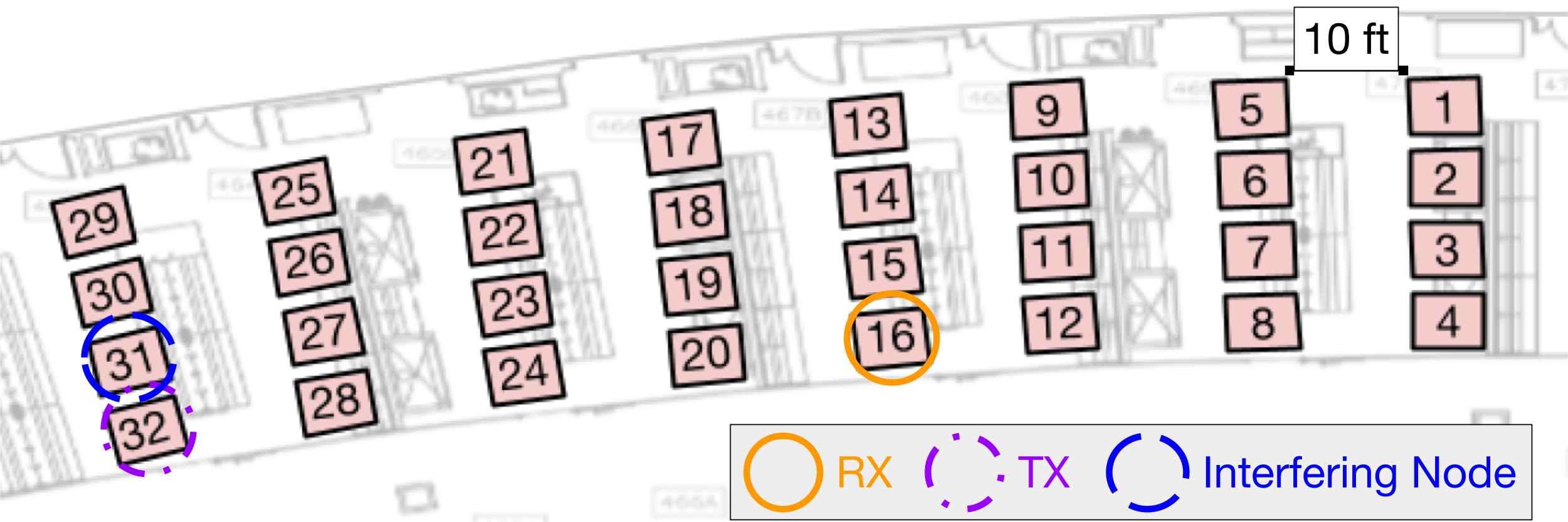}
    \caption{The experimental setup on the Arena testbed.}
    \label{fig:arena}
\end{figure}

We collect data and prototype \textit{DeepSweep} over the air on the Arena testbed~\cite{bertizzolo_arena_2020}, a 64-antenna wireless testbed with 24 software-defined radios (SDRs). The setup is illustrated in Fig.~\ref{fig:arena}. We consider an OFDM system (WiFi in this case) operating on the 2.4GHz ISM band. The setup includes a WiFi transmitter and receiver pair as well as a narrowband interfering node, all implemented using NI USRP X310 SDR radios. WiFi nodes transmit over a 10~MHz WiFi channel while the interfering signal consists of a narrowband signal occupying 156~KHz, equal to 1.5\% of the WiFi signal bandwidth, i.e., the size of a single OFDM subcarrier. The location of the interfering signal changes throughout data collection and testing experiments and selected among 8 possible locations in frequency.

\textit{DeepSweep}'s components are implemented on top of the GNU Radio IEEE 802.11 a/g/p open-source stack~\cite{bloessl2013ieee}, while the CNN is implemented in TensorFlow and interfaced with the IQs received from the RF frontend via GNU Radio.


\subsection{Over-the-air dataset}
\label{subsec:fft_set}

As mentioned earlier, \textit{DeepSweep} has been trained on data exclusively collected over the air on the Arena testbed. The dataset used to train the spectrum sensing CNN consists of more than 400,000 unique entries representing the collected frequency-domain IQ samples. Specifically, the data collection phase involves the collection of $N_{\mathrm{TIME}}=1024$ IQ samples in the time domain, which are then projected into the frequency domain via the FFT operation (as illustrated in Fig.~\ref{fig:deepsweep}). In our data collection, we have used an FFT size of $N_{\mathrm{FFT}}=256$. Recall that the 10~MHz WiFi channel consists of 64 OFDM subcarriers (only 52 are used for data and pilot transmission). Therefore, each OFDM subcarrier is represented by 4 IQ samples in the frequency domain. 
As discussed in Section \ref{subsec:usecase}, in this paper we focus on detecting narrowband interference. To generate a labeled dataset, the interfering radio shown in Fig. \ref{fig:arena} is configured to programmatically generate interference on 8 specific subcarriers. For each collection cycle, we collect IQ samples by fixing the subcarrier being affected by interference and eventually attach a label that specifies the subcarrier. 

In order to account for different SNR and channel conditions, we also collect data across different days and by using different transmission gains of the radios to generate a dataset that is as general as possible and better describes real-world conditions. 





\section{Experimental Results}
\label{sec:results}

In this section, we evaluate \textit{DeepSweep} performance and highlight how processing separate portions of spectrum independently with high-resolution lightweight models can improve training and inference times while delivering high accuracy. For this reason, we compare our model against VGG16~\cite{simonyan2014very}. We chose this specific CNN due to the fact previous research has shown this model is a strong candidate when creating spectrum sensing techniques, where in \cite{vgg16_spec_sense}, a model using VGG-16 reached accuracies up to 93\%. However, where this model excels in accuracy, the latency values are much slower than other models due to the number of convolutions required. Since this is a popular model, we wanted to show a highly accurate model with less added latency. 
In the following, we present results by considering $G\in\{1,2,4,8\}$, which corresponds to splitting the input to \textit{DeepSweep} into groups each covering a corresponding bandwidth of $\{10,5,2.5,1.25\}$~MHz, respectively. For each of these values, we have trained a different CNN (both the one we have proposed in Section \ref{subsec:cnn} as well as VGG16). 



\subsection{Training time and accuracy}
\label{subsec:training}

\begin{figure}[!t]
    \centering
    \includegraphics[width=0.85\linewidth]{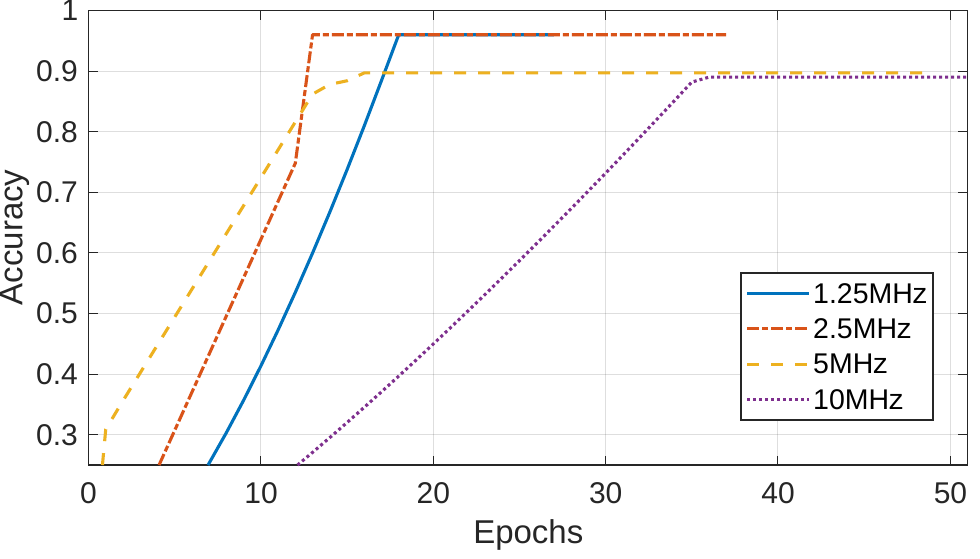}
    \caption{Validation accuracy over the model training for the \textit{DeepSweep} CNN.}
    \label{fig:val_acc}
\end{figure}

\begin{figure}[!b]
    \vspace{-5pt} 
    \centering
    \includegraphics[width=.85\linewidth]{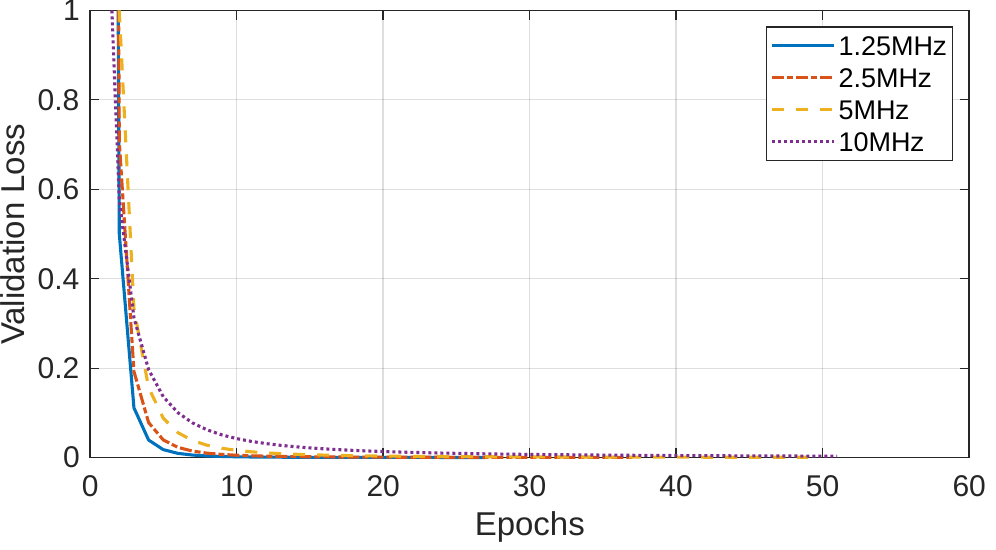}
    \caption{Validation loss over the model training for the \textit{DeepSweep} CNN.}
    \label{fig:val_loss}
\end{figure}

In Fig.~\ref{fig:val_acc}, we show the validation accuracy, and in Fig.~\ref{fig:val_loss} we show the validation loss over the training epochs for our model and for different input sizes (in this case shown as the corresponding bandwidth effectively monitored by the CNN). 

In general, the models with the input size of 1.25~MHz and 2.5~MHz are the fastest to be trained and converge in less than 20 epochs. They also deliver the highest accuracy.
As expected, the slowest model has an input size of 10~MHz and also achieves a lower accuracy overall.

\begin{figure}[!t]
    \centering
    \includegraphics[width=.95\linewidth]{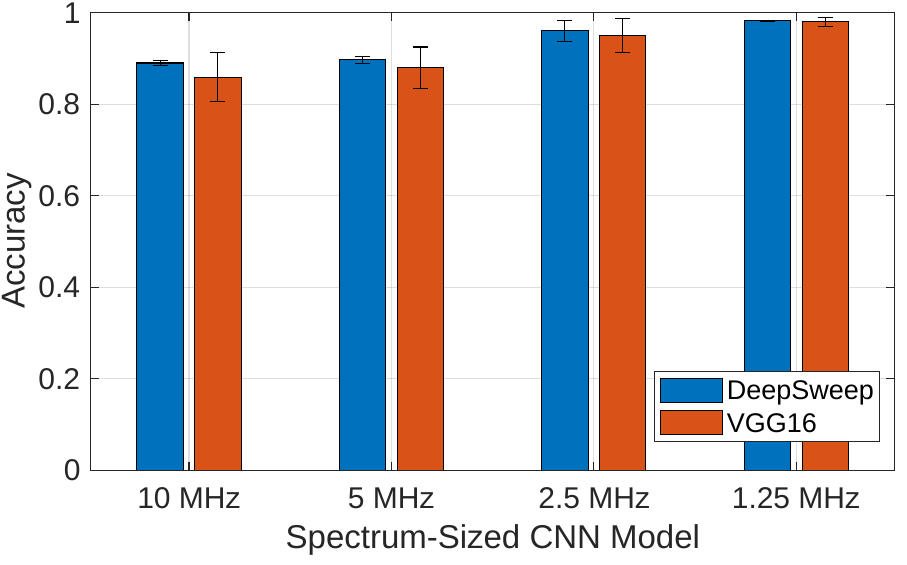}
    \caption{Testing accuracy comparison between \textit{DeepSweep} and VGG16 for different input bandwidth size.}
    \label{fig:acc_comp}
    \vspace{-10pt}
\end{figure}

\subsection{Testing accuracy}
\label{subsec:acc}

The results presented in this section have been obtained by testing our spectrum sensing CNN on completely unseen data that was collected on a different day and used to train the CNN (see Section \ref{subsec:fft_set}). The goal is to show that \textit{DeepSweep} can effectively generalize across multiple days.

In Fig.~\ref{fig:acc_comp}, we show the testing accuracy for \textit{DeepSweep} and VGG16 models as a function of the input bandwidth size. We notice that the testing accuracy increases as we reduce the input bandwidth size, meaning that both CNNs can better identify narrowband interference when they observe a smaller portion of the WiFi bandwidth only. Indeed, observing a much smaller portion of the spectrum at a time reduces the likelihood of making a misclassification, which instead is more likely to happen when processing the entire 10~MHz band. When considering a 10~MHz input, \textit{DeepSweep} is more accurate than VGG16 by an additional 3\%, while this gap reduces to 0.2\% when we consider an input representing 1.25~MHz only. 

\begin{figure}[!b]
    \vspace{-10pt}
    \centering
    \includegraphics[width=0.85\linewidth]{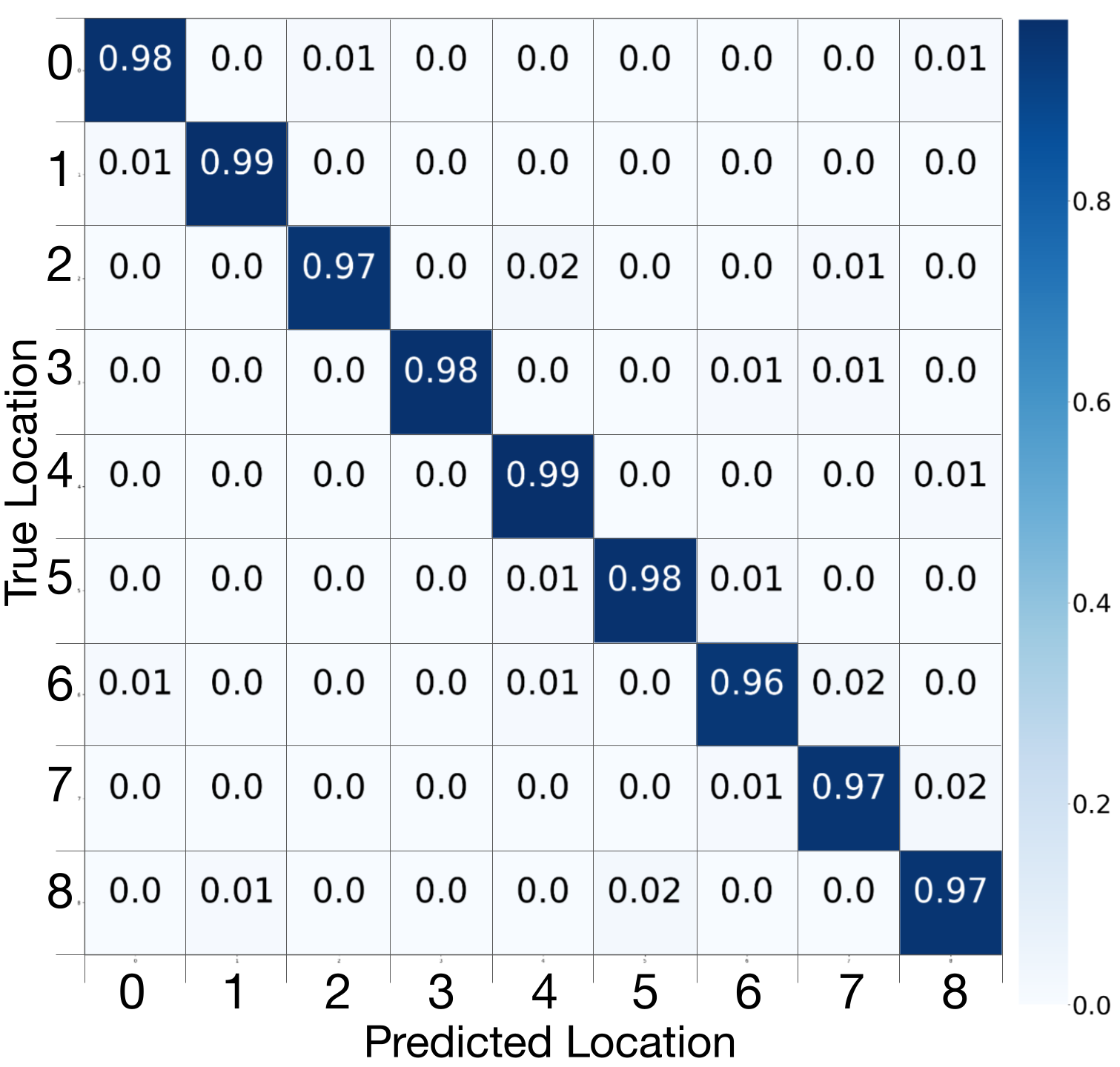}
    \caption{Confusion matrix when interference can appear in 8 possible locations.}
    \label{fig:jam_cms}
\end{figure}

Another important aspect of detecting interference is being able to accurately locate it to determine portions of the spectrum being affected. 
Fig. \ref{fig:jam_cms} shows the accuracy in locating interference at different locations. We consider the case where we monitor 1.25~MHz only, i.e., the monitored portion of spectrum contains OFDM 8 subcarriers, and the jammer can jam one subcarrier at a time. In this case, \textit{DeepSweep} must be able to detect which subcarrier is being attacked.
The overall accuracy in locating interference is 98\%, showing that \textit{DeepSweep} can accurately detect and locate interference. 

\subsection{Inference time}
\label{subsec:lat}



\begin{figure}[!t]
    \centering
    \includegraphics[width=.85\linewidth]{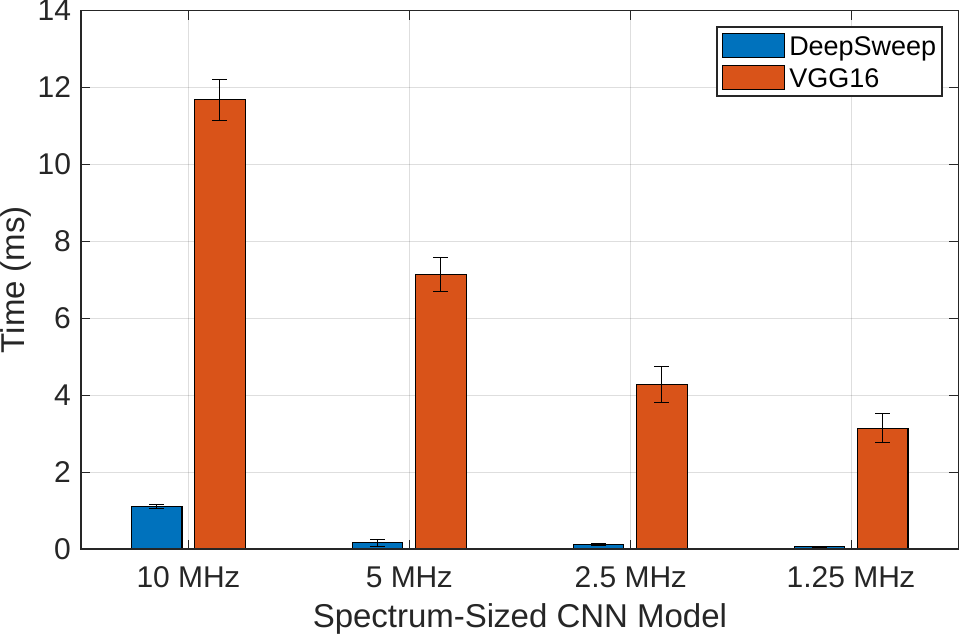}
    \caption{Latency values of \textit{DeepSweep} and VGG16 models based on input bandwidth size deployed in an over-the-air environment.}
    \label{fig:lat_val}
    \vspace{-9pt}
\end{figure}
Fig.~\ref{fig:lat_val} shows the inference time of both \textit{DeepSweep} and VGG16 when deployed to perform real-time inference on our testbed. 
Our results show that our model performs inference significantly faster than VGG16. For an input size of 10~MHz, our model outperforms the VGG16 model by being in the worst case 10$\times$ faster than VGG16. The improvement in speed becomes more and more evident (up to 38 times faster) when we process chunks of $1.25$~MHz of bandwidth in parallel. 
Together with Fig. \ref{fig:acc_comp}, these results show that our model can significantly reduce inference time without sacrificing any accuracy.

To demonstrate that \textit{DeepSweep} can operate in real-time, it needs to process IQ samples timely and as they arrive at the receiver. We know that the amount of time to collect $N_{\mathrm{TIME}}$ IQs in the time domain at sampling rate $f_s$ is $ T = N_{\mathrm{TIME}} \times \frac{1}{f_s}$. In our case, $f_S = 10$~MHz and $T=0.1024ms$.
From Fig.~\ref{fig:lat_val}, we know that \textit{DeepSweep} processes an input of 1.25~MHz in 0.055~ms, which is actually 46\% faster than the time required to collect the amount of IQ samples. 



\section{Conclusion}
\label{sec:conclusion}

In this paper, we have proposed \textit{DeepSweep}, a novel data-driven transceiver design that enables scalable, accurate, and fast spectrum sensing without interrupting ongoing demodulation and decoding operations. Our approach is based on a spectrum sensing pipeline that splits incoming waveforms into equally sized spectrum chunks and processes them in parallel and in batches by a high-resolution deep learning model. Through our experiments, we have shown that our approach can reduce training and inference times by more than 2$\times$ and 10$\times$ respectively, compared to a model with the same resolution processing the entire monitored band as a whole. Furthermore, we have demonstrated the effectiveness of \textit{DeepSweep} in a real-world scenario, where we aimed to detect and locate narrowband interfering signals with high accuracy. Our experiments using COTS wireless radios and over-the-air data show that \textit{DeepSweep} achieves up to 98\% detection accuracy and has an inference latency lower than 1~ms, indicating that \textit{DeepSweep} is a promising solution for a variety of spectrum sensing tasks. Future work will focus on demonstrating our approach in different application scenarios including waveform classification and spectrum hole detection.





\footnotesize
\balance
\bibliographystyle{IEEEtran}
\bibliography{bib}

\end{document}